\documentclass[%
 reprint,
superscriptaddress,
 amsmath,amssymb,
 aps,
]{revtex4-2}

\usepackage{graphicx}
\usepackage{dcolumn}
\usepackage{bm}
\usepackage[mathlines]{lineno}
\usepackage{verbatim}
\usepackage{xcolor}
\RequirePackage{xspace}
\RequirePackage{wasysym}

\newcommand{\onbb}{$0\nu\beta\beta$\xspace}
\newcommand{\nnbb}{$2\nu\beta\beta$\xspace}

\newcommand{\qbb}{$Q_{\beta\beta}$\xspace}

\newcommand{\Mo}{$^{100}$Mo\xspace}

\newcommand{\lmo}{LMO\xspace}
\newcommand{\enrlmo}{Li$_{2}${}$^{100}$MoO$_4$\xspace}

\newcommand\edit[1]{\textcolor{black}{#1}}

\newcommand{\Tl}{$^{208}\mathrm{Tl}$\xspace}

\newcommand{\Co}{$^{60}\mathrm{Co}$\xspace}
\newcommand{\K}{$^{40}\mathrm{K}$\xspace}

\newcommand{\Po}{$^{210}\mathrm{Po}$\xspace}

\newcommand{\al}{$\alpha$\xspace}

\newcommand{\ga}{$\gamma$\xspace}
\newcommand{\gabe}{$\gamma$\&$\beta$\xspace}

\newcommand{\ky}{kg$\times$yr\xspace}
\newcommand{\ckky}{counts/(keV$\times$kg$\times$yr)\xspace}

\begin{document}

\preprint{APS/123-QED}

\title{New Limit for Neutrinoless Double-Beta Decay of $^{100}$Mo from the CUPID-Mo Experiment}
\collaboration{CUPID-Mo Collaboration}

\author{E.~Armengaud }
\affiliation{IRFU, CEA, Universit\'{e} Paris-Saclay, F-91191 Gif-sur-Yvette, France }

\author{C.~Augier }
\affiliation{Univ Lyon, Universit\'{e} Lyon 1, CNRS/IN2P3, IP2I-Lyon, F-69622, Villeurbanne, France }

\author{A.~S.~Barabash }
\affiliation{National Research Centre Kurchatov Institute, Institute of Theoretical and Experimental Physics, 117218 Moscow, Russia }

\author{F.~Bellini }
\affiliation{Dipartimento di Fisica, Sapienza Universit\`a di Roma, P.le Aldo Moro 2, I-00185, Rome, Italy }
\affiliation{INFN, Sezione di Roma, P.le Aldo Moro 2, I-00185, Rome, Italy}

\author{G.~Benato }
\affiliation{INFN, Laboratori Nazionali del Gran Sasso, I-67100 Assergi (AQ), Italy }

\author{A.~Beno\^{\i}t }
\affiliation{CNRS-N\'eel, 38042 Grenoble Cedex 9, France }

\author{M.~Beretta }
\affiliation{ University of California, Berkeley, California 94720, USA }

\author{L.~Berg\'e }
\affiliation{Universit\'e Paris-Saclay, CNRS/IN2P3, IJCLab, 91405 Orsay, France }

\author{J.~Billard }
\affiliation{Univ Lyon, Universit\'{e} Lyon 1, CNRS/IN2P3, IP2I-Lyon, F-69622, Villeurbanne, France }

\author{Yu.~A.~Borovlev }
\affiliation{Nikolaev Institute of Inorganic Chemistry, 630090 Novosibirsk, Russia }

\author{Ch.~Bourgeois } 
\affiliation{Universit\'e Paris-Saclay, CNRS/IN2P3, IJCLab, 91405 Orsay, France }

\author{V.~B.~Brudanin } \thanks{Deceased} 
\affiliation{Laboratory of Nuclear Problems, JINR, 141980 Dubna, Moscow region, Russia }

\author{P.~Camus }
\affiliation{CNRS-N\'eel, 38042 Grenoble Cedex 9, France }

\author{L.~Cardani }
\affiliation{INFN, Sezione di Roma, P.le Aldo Moro 2, I-00185, Rome, Italy}

\author{N.~Casali }
\affiliation{INFN, Sezione di Roma, P.le Aldo Moro 2, I-00185, Rome, Italy}

\author{A.~Cazes }
\affiliation{Univ Lyon, Universit\'{e} Lyon 1, CNRS/IN2P3, IP2I-Lyon, F-69622, Villeurbanne, France }

\author{M.~Chapellier }
\affiliation{Universit\'e Paris-Saclay, CNRS/IN2P3, IJCLab, 91405 Orsay, France }

\author{F.~Charlieux }
\affiliation{Univ Lyon, Universit\'{e} Lyon 1, CNRS/IN2P3, IP2I-Lyon, F-69622, Villeurbanne, France }

\author{D.~Chiesa}
\affiliation{Dipartimento di Fisica, Universit\`{a} di Milano-Bicocca, I-20126 Milano, Italy }
\affiliation{INFN, Sezione di Milano-Bicocca, I-20126 Milano, Italy}

\author{M.~de~Combarieu }
\affiliation{IRAMIS, CEA, Universit\'{e} Paris-Saclay, F-91191 Gif-sur-Yvette, France }

\author{I.~Dafinei }
\affiliation{INFN, Sezione di Roma, P.le Aldo Moro 2, I-00185, Rome, Italy}

\author{F.~A.~Danevich }
\affiliation{Institute for Nuclear Research of NASU, 03028 Kyiv, Ukraine }

\author{M.~De~Jesus }
\affiliation{Univ Lyon, Universit\'{e} Lyon 1, CNRS/IN2P3, IP2I-Lyon, F-69622, Villeurbanne, France }

\author{T.~Dixon}
\affiliation{ University of California, Berkeley, California 94720, USA }

\author{L.~Dumoulin }
\affiliation{Universit\'e Paris-Saclay, CNRS/IN2P3, IJCLab, 91405 Orsay, France }

\author{K.~Eitel }
\affiliation{Karlsruhe Institute of Technology, Institute for Astroparticle Physics, 76021 Karlsruhe, Germany }

\author{F.~Ferri }
\affiliation{IRFU, CEA, Universit\'{e} Paris-Saclay, F-91191 Gif-sur-Yvette, France }

\author{B.~K.~Fujikawa }
\affiliation{ Lawrence Berkeley National Laboratory, Berkeley, California 94720, USA }

\author{J.~Gascon }
\affiliation{Univ Lyon, Universit\'{e} Lyon 1, CNRS/IN2P3, IP2I-Lyon, F-69622, Villeurbanne, France }

\author{L.~Gironi }
\affiliation{Dipartimento di Fisica, Universit\`{a} di Milano-Bicocca, I-20126 Milano, Italy }
\affiliation{INFN, Sezione di Milano-Bicocca, I-20126 Milano, Italy}

\author{A.~Giuliani}\email{e-mail: andrea.giuliani@ijclab.in2p3.fr} 
\affiliation{Universit\'e Paris-Saclay, CNRS/IN2P3, IJCLab, 91405 Orsay, France }

\author{V.~D.~Grigorieva }
\affiliation{Nikolaev Institute of Inorganic Chemistry, 630090 Novosibirsk, Russia }

\author{M.~Gros }
\affiliation{IRFU, CEA, Universit\'{e} Paris-Saclay, F-91191 Gif-sur-Yvette, France }

\author{E.~Guerard}
\affiliation{Universit\'e Paris-Saclay, CNRS/IN2P3, IJCLab, 91405 Orsay, France }

\author{D.~L.~Helis }
\affiliation{IRFU, CEA, Universit\'{e} Paris-Saclay, F-91191 Gif-sur-Yvette, France }

\author{H.~Z.~Huang }
\affiliation{Key Laboratory of Nuclear Physics and Ion-beam Application (MOE), Fudan University, Shanghai 200433, PR China }

\author{R.~Huang }
\affiliation{ University of California, Berkeley, California 94720, USA }

\author{J.~Johnston }
\affiliation{Massachusetts Institute of Technology, Cambridge, MA 02139, USA }

\author{A.~Juillard }
\affiliation{Univ Lyon, Universit\'{e} Lyon 1, CNRS/IN2P3, IP2I-Lyon, F-69622, Villeurbanne, France }

\author{H.~Khalife }
\affiliation{Universit\'e Paris-Saclay, CNRS/IN2P3, IJCLab, 91405 Orsay, France }

\author{M.~Kleifges }
\affiliation{Karlsruhe Institute of Technology, Institute for Data Processing and Electronics, 76021 Karlsruhe, Germany }

\author{V.~V.~Kobychev }
\affiliation{Institute for Nuclear Research of NASU, 03028 Kyiv, Ukraine }

\author{Yu.~G.~Kolomensky }
\affiliation{ University of California, Berkeley, California 94720, USA }
\affiliation{ Lawrence Berkeley National Laboratory, Berkeley, California 94720, USA }%

\author{S.I.~Konovalov }
\affiliation{National Research Centre Kurchatov Institute, Institute of Theoretical and Experimental Physics, 117218 Moscow, Russia }

\author{A.~Leder }
\affiliation{Massachusetts Institute of Technology, Cambridge, MA 02139, USA }

\author{P.~Loaiza }
\affiliation{Universit\'e Paris-Saclay, CNRS/IN2P3, IJCLab, 91405 Orsay, France }

\author{L.~Ma }
\affiliation{Key Laboratory of Nuclear Physics and Ion-beam Application (MOE), Fudan University, Shanghai 200433, PR China }

\author{E.~P.~Makarov }
\affiliation{Nikolaev Institute of Inorganic Chemistry, 630090 Novosibirsk, Russia }

\author{P.~de~Marcillac }
\affiliation{Universit\'e Paris-Saclay, CNRS/IN2P3, IJCLab, 91405 Orsay, France }

\author{R.~Mariam}
\affiliation{Universit\'e Paris-Saclay, CNRS/IN2P3, IJCLab, 91405 Orsay, France }

\author{L.~Marini }
\affiliation{ University of California, Berkeley, California 94720, USA }
\affiliation{ Lawrence Berkeley National Laboratory, Berkeley, California 94720, USA }
\affiliation{INFN, Laboratori Nazionali del Gran Sasso, I-67100 Assergi (AQ), Italy }

\author{S.~Marnieros }
\affiliation{Universit\'e Paris-Saclay, CNRS/IN2P3, IJCLab, 91405 Orsay, France }

\author{D.~Misiak }
\affiliation{Univ Lyon, Universit\'{e} Lyon 1, CNRS/IN2P3, IP2I-Lyon, F-69622, Villeurbanne, France }

\author{X.-F.~Navick }
\affiliation{IRFU, CEA, Universit\'{e} Paris-Saclay, F-91191 Gif-sur-Yvette, France }

\author{C.~Nones }
\affiliation{IRFU, CEA, Universit\'{e} Paris-Saclay, F-91191 Gif-sur-Yvette, France }

\author{E.~B.~Norman}
\affiliation{ University of California, Berkeley, California 94720, USA }

\author{V.~Novati } \thanks{Now at: Northwestern University, Evanston, IL 60208, USA } 
\affiliation{Universit\'e Paris-Saclay, CNRS/IN2P3, IJCLab, 91405 Orsay, France }

\author{E.~Olivieri }
\affiliation{Universit\'e Paris-Saclay, CNRS/IN2P3, IJCLab, 91405 Orsay, France }

\author{J.~L.~Ouellet }
\affiliation{Massachusetts Institute of Technology, Cambridge, MA 02139, USA }

\author{L.~Pagnanini }
\affiliation{INFN, Gran Sasso Science Institute, I-67100 L'Aquila, Italy}
\affiliation{INFN, Laboratori Nazionali del Gran Sasso, I-67100 Assergi (AQ), Italy }

\author{P.~Pari }
\affiliation{IRAMIS, CEA, Universit\'{e} Paris-Saclay, F-91191 Gif-sur-Yvette, France }

\author{L.~Pattavina }
\affiliation{INFN, Laboratori Nazionali del Gran Sasso, I-67100 Assergi (AQ), Italy }
\affiliation{Physik Department, Technische Universit\"at M\"unchen, Garching D-85748, Germany }

\author{B.~Paul }
\affiliation{IRFU, CEA, Universit\'{e} Paris-Saclay, F-91191 Gif-sur-Yvette, France }

\author{M.~Pavan }
\affiliation{Dipartimento di Fisica, Universit\`{a} di Milano-Bicocca, I-20126 Milano, Italy }
\affiliation{INFN, Sezione di Milano-Bicocca, I-20126 Milano, Italy}

\author{H.~Peng }
\affiliation{Department of Modern Physics, University of Science and Technology of China, Hefei 230027, PR China }

\author{G.~Pessina }
\affiliation{INFN, Sezione di Milano-Bicocca, I-20126 Milano, Italy}

\author{S.~Pirro }
\affiliation{INFN, Laboratori Nazionali del Gran Sasso, I-67100 Assergi (AQ), Italy }

\author{D.~V.~Poda }
\affiliation{Universit\'e Paris-Saclay, CNRS/IN2P3, IJCLab, 91405 Orsay, France }

\author{O.~G.~Polischuk }
\affiliation{Institute for Nuclear Research of NASU, 03028 Kyiv, Ukraine }

\author{S.~Pozzi }
\affiliation{INFN, Sezione di Milano-Bicocca, I-20126 Milano, Italy}

\author{E.~Previtali }
\affiliation{Dipartimento di Fisica, Universit\`{a} di Milano-Bicocca, I-20126 Milano, Italy }
\affiliation{INFN, Sezione di Milano-Bicocca, I-20126 Milano, Italy}

\author{Th.~Redon }
\affiliation{Universit\'e Paris-Saclay, CNRS/IN2P3, IJCLab, 91405 Orsay, France }

\author{A.~Rojas }
\affiliation{LSM, Laboratoire Souterrain de Modane, 73500 Modane, France }

\author{S.~Rozov }
\affiliation{Laboratory of Nuclear Problems, JINR, 141980 Dubna, Moscow region, Russia }

\author{C.~Rusconi }
\affiliation{Department of Physics and Astronomy, University of South Carolina, SC 29208, Columbia, USA }

\author{V.~Sanglard }
\affiliation{Univ Lyon, Universit\'{e} Lyon 1, CNRS/IN2P3, IP2I-Lyon, F-69622, Villeurbanne, France }

\author{J.~A.~Scarpaci}
\affiliation{Universit\'e Paris-Saclay, CNRS/IN2P3, IJCLab, 91405 Orsay, France }

\author{K.~Sch\"affner }
\affiliation{INFN, Laboratori Nazionali del Gran Sasso, I-67100 Assergi (AQ), Italy }

\author{B.~Schmidt } \thanks{Now at: Northwestern University, Evanston, IL 60208, USA }
\affiliation{ Lawrence Berkeley National Laboratory, Berkeley, California 94720, USA }

\author{Y.~Shen }
\affiliation{Key Laboratory of Nuclear Physics and Ion-beam Application (MOE), Fudan University, Shanghai 200433, PR China }

\author{V.~N.~Shlegel }
\affiliation{Nikolaev Institute of Inorganic Chemistry, 630090 Novosibirsk, Russia }

\author{B.~Siebenborn }
\affiliation{Karlsruhe Institute of Technology, Institute for Astroparticle Physics, 76021 Karlsruhe, Germany }

\author{V.~Singh }
\affiliation{ University of California, Berkeley, California 94720, USA }

\author{C.~Tomei }
\affiliation{INFN, Sezione di Roma, P.le Aldo Moro 2, I-00185, Rome, Italy}

\author{V.~I.~Tretyak }
\affiliation{Institute for Nuclear Research of NASU, 03028 Kyiv, Ukraine }

\author{V.~I.~Umatov }
\affiliation{National Research Centre Kurchatov Institute, Institute of Theoretical and Experimental Physics, 117218 Moscow, Russia }

\author{L.~Vagneron }
\affiliation{Univ Lyon, Universit\'{e} Lyon 1, CNRS/IN2P3, IP2I-Lyon, F-69622, Villeurbanne, France }

\author{M.~Vel\'azquez }
\affiliation{Universit\'e Grenoble Alpes, CNRS, Grenoble INP, SIMAP, 38402 Saint Martin d'H\'eres, France }

\author{B.~Welliver }
\affiliation{ Lawrence Berkeley National Laboratory, Berkeley, California 94720, USA }

\author{L.~Winslow }
\affiliation{Massachusetts Institute of Technology, Cambridge, MA 02139, USA }

\author{M.~Xue }
\affiliation{Department of Modern Physics, University of Science and Technology of China, Hefei 230027, PR China }

\author{E.~Yakushev }
\affiliation{Laboratory of Nuclear Problems, JINR, 141980 Dubna, Moscow region, Russia }

\author{M.~Zarytskyy}
\affiliation{Institute for Nuclear Research of NASU, 03028 Kyiv, Ukraine }

\author{A.~S.~Zolotarova }
\affiliation{Universit\'e Paris-Saclay, CNRS/IN2P3, IJCLab, 91405 Orsay, France }

\date{\today}

\begin{abstract}

The CUPID-Mo experiment at the Laboratoire Souterrain de Modane (France) is a demonstrator for CUPID, the next-generation ton-scale bolometric \onbb experiment. It consists of a 4.2\,kg array of 20 enriched \enrlmo\ scintillating bolometers to search for the lepton number violating process of \onbb\ decay in \Mo. 
With more than one year of operation (\Mo exposure of 1.17 \ky for physics data), no event in the region of interest and hence no evidence for \onbb is observed. 
We report a new limit on the half-life of \onbb decay in \Mo of $T_{1/2}  > 1.5 \times 10^{24}\,$yr at 90\% C.I. The limit corresponds to an effective Majorana neutrino mass $\langle m_{\beta\beta} \rangle$\ $<$ (0.31--0.54)$\,$eV, dependent on the nuclear matrix element in the light Majorana neutrino exchange interpretation. 
\end{abstract}

\maketitle


The discovery that neutrinos are massive particles through the evidence of neutrino flavor oscillations \cite{Esteban:2019} opens the question of neutrino mass generation. Instead of having Dirac nature as charged leptons and quarks, the scale of neutrino masses could be well motivated by the Majorana theory \cite{Mohapatra:1980a,Schechter:1980a}. In this scenario neutrinos could coincide with their antimatter partner~\cite{Majorana:1937a, Racah:1937} which would have a tremendous impact on our vision of Nature, implying the violation of the total lepton number $L$ as well as for the matter-antimatter asymmetry in the Universe~\cite{Fukugita:1986,Davidson:2008a}.

The distinction between Dirac and Majorana behavior is an extreme experimental challenge. Neutrinoless double-beta  (\onbb) decay is the traditional and the most sensitive tool to probe the Majorana nature of neutrinos. This process is a nuclear transition consisting in the transformation of an even-even nucleus into a lighter isobar containing two more protons and accompanied by the emission of two electrons and no other particles, with a change of the lepton number $L$ by two units~\cite{Furry:1939,Bilenky:2015,DellOro:2016tmg,Dolinski:2019a}. An observation of this hypothetical process would establish that neutrinos are Majorana particles~\cite{Schechter:1982}. The current most stringent limits on \onbb decay half-lives are at the level of $ 10^{25}$--$10^{26}$ yr in $^{136}$Xe, $^{76}$Ge and $^{130}$Te \cite{Gando:2016, Anton:2019, Alvis:2019, Agostini:2020, Adams2020}.
\onbb decay can be induced by a variety of mechanisms~\cite{Dolinski:2019a,Bilenky:2015,Deppisch:2012,Rodejohann:2012}. Among them, the so-called mass mechanism --- consisting in the exchange of a virtual light Majorana neutrino --- represents a minimal extension of the Standard Model. In this mechanism, the \onbb decay rate is proportional to the square of the effective Majorana neutrino mass $\langle m_{\beta\beta} \rangle$, a linear combination of the three neutrino mass eigenvalues which fixes the absolute neutrino mass scale. Present limits on $\langle m_{\beta\beta} \rangle$ are in the range of $(0.06$--$0.6)$ eV~\cite{Dolinski:2019a}, assuming that the axial charge $g_A$ is not quenched and equal to the free nucleon value of $\simeq 1.27$  \cite{Suhonen:2019,Gysbers:2019,Simkovic:2018b}.

The distinctive signal of \onbb decay is a peak in the sum energy spectrum of the two emitted electrons at the total available energy \qbb of the \onbb transition.
Among the 35 natural double-beta emitters (\onbb candidate isotopes)~\cite{Tretyak:2002}, only a few of them are experimentally relevant. These favorable candidates feature a high \qbb ($> 2$~MeV), which leads to a high decay probability and to a low background level in the signal region. At the same time, these candidates exhibit a high natural abundance of the isotope of interest and/or a technically feasible isotopic enrichment at the tonne scale.

Low-temperature calorimeters, often named bolometers, are the detectors of choice for several experimental efforts, including the one reported here. Featuring high energy resolution, high efficiency, and flexibility in detector-material choice~\cite{Fiorini:1984,Giuliani:2012,Poda:2017}, bolometers are perfectly tailored to \onbb search. These detectors consist in a single crystal that contains the \onbb source coupled to a temperature sensor. The signal is collected at very low temperatures $\lesssim 20$\,mK for large (0.1--1\,kg) bolometers and consists of a thermal pulse registered by the sensor. 

A detector embedding a candidate with ${Q_{\beta\beta} > 2615\,}$keV is an optimal choice in terms of background control, as the bulk of the $\gamma$ natural radioactivity ends at 2615\,keV, corresponding to the energy of the \Tl line in the $^{232}$Th decay chain. However, the energy region above $\sim 2.6\,$MeV is dominated by events due to surface radioactive contamination, especially energy-degraded $\alpha$~particles~\cite{Alduino:2018,Adams2020}, as shown by the results of  CUORE, the largest \onbb bolometric experiment currently under way. 

A dual readout of light --- scintillation or Cherenkov --- in addition to the thermal signal allows for the discrimination of $\alpha$ events in various targets~\cite{Pirro:2005ar,Poda:2017,Tabarelli:2010,Giuliani:2012,Giuliani:2018, Azzolini:2019nmi,Azzolini:2019tta,  Alenkov:2019}. 
This technology has been developed for the scintillating \enrlmo crystals used in CUPID-Mo by the LUMINEU Collaboration \cite{Armengaud:2017,Poda:2017a} and its effectiveness is described together with the experimental setup in  \cite{Armengaud:2020}. 
The isotope of interest \Mo features a $Q_{\beta\beta}$  of $ (3034.40 \pm 0.17) $\,keV \cite{Rahaman:2008} and a natural abundance of 9.7\% making large-scale enrichment viable by gas centrifuge isotopic separation \cite{CUPIDInterestGroup:2019inu}. In CUPID-Mo, it is embedded into \enrlmo (\lmo) crystals by a double low-thermal-gradient Czochralski crystallization process~\cite{Grigorieva:2017} from enriched Mo previously used in the NEMO-3 experiment \cite{Arnold:2015}.  
A total of twenty cylindrical $\sim210\,$g crystals are stacked into 5 towers which results in a \Mo mass of $ (2.258\pm0.005)\,$kg with an average \Mo isotopic abundance of $(96.6\pm0.2)\%$. 
Round Ge wafers, attached to the bottom of each \lmo detector, are used as bolometric light detectors (LDs). Due to the stacking into the 4-layer tower most \lmo detectors have a direct line of sight to a LD both at the top and bottom, except for the top crystal of each tower which has a Cu lid on one side \cite{Armengaud:2020}.
The \lmo crystals as well as the LDs are instrumented with Neutron-Transmutation-Doped (NTD)-Ge sensors \cite{Haller:1994}. The towers are installed with a mechanical decoupling inside the EDELWEISS cryogenic infrastructure~\cite{Armengaud:2017b, Hehn:2016} at the Laboratoire Souterrain de Modane in France. 


The data of the present analysis have been acquired over a 380 day period  between March 2019 and April 2020 at operation temperatures of 20.7 and 22\,mK. About $82\,\% $ of the time was devoted to the \onbb search, split into 240 days of physics data and 73 days of calibration data. 
The physics data is grouped into a total of 10 data-sets with consistent operation conditions. In the following we  consider 213 out of the 240 days of physics data in 7 (1--2 month long) data-sets and reject 3 ($\sim$1~week long) data-sets due to their small associated calibration statistics.
From these 7 data-sets we exclude periods of temperature instabilities, disturbances in the underground laboratory and periods of excessive noise on the individual detectors reducing the physics exposure by 6\%. We reject one of the twenty \lmo bolometers that shows an abnormal performance \cite{Armengaud:2020}
and obtain a physics exposure of 2.16 \ky (\enrlmo).


All data are acquired as a continuous stream with $500 \,$Hz sampling frequency and analyzed with a software package developed by the CUORE \cite{Alduino:2016} and CUPID-0 \cite{Azzolini2018b} Collaborations, first used in CUPID-Mo in \cite{Armengaud:2020, Schmidt:2020}. 
\edit{We estimate pulse amplitudes with an optimum filter \cite{Gatti:1986}, designed to maximise the signal to noise ratio for a known signal and noise spectrum with 3\,s long pulse traces for both the \lmo and LD channels. The data were triggered offline using the optimum filter \cite{DiDomizio:2010ph, Adams2020} obtaining 90\% trigger efficiency at typical (median) energies of 9.4\,keV / 0.5\,keV for the LMOs / LDs. 
The LMOs analysis and coincidence thresholds have been set at 45\,keV, well above this efficiency turn on. For each signal on an LMO detector we evaluate the resolution weighted average light signal of the two (one) adjacent LDs to discriminate $\alpha$ events, exhibiting $\sim$20\% of the light yield of \gabe events of the same energy \cite{Armengaud:2020}. 
We calibrate the response of the \lmo detectors with a 2nd-order polynomial using the four labeled peaks from the U/Th calibration data shown in Fig. 1 in red (see \cite{Schmidt:2020}) and cross-calibrate the LD against the LMO signals.}
We confirm the LMO's energy scale in background data fitting a 2nd-order polynomial  in reconstructed-to-expected peak position of the 352, 583, 609, 1461 and 2615\,keV 
peaks and observe no systematic deviation. The extrapolation for the position of \qbb agrees to within  ${E_{bias}^{Q_{\beta\beta}} = (-0.2\pm0.4)\,\mathrm{keV}}$.

We adopt a blinding strategy removing all events in a $\pm 50\,$keV window around \qbb to avoid any bias in the optimization of our analysis procedures and consider the following event selections. For events (i) to be contained in a single crystal and in anti-coincidence with a triple module trigger and energy deposit in the muon-veto system \cite{Schmidt:2013} based on a $\pm$100\,ms time window; (ii) to have a single trigger in each 3-s pulse window; (iii) to have a flat pre-trace with a slope of less than 15 median absolute deviations; (iv) to have a pulse shape compatible with the principal components (PC) established by a newly developed PC-Analysis described in \cite{Huang:2020}. This cut is optimized using calibration data by maximizing a hypothetical discovery sensitivity for a \onbb process equal in half-live to the previous best limit \cite{Arnold:2015}; (v) to have the expected light yield for \gabe events and no difference in top and bottom LDs. Both of these cuts are set to obtain close to full coverage at $\pm 3\,\sigma$ each, based on a Gaussian fit of the light yield in calibration data. The energy dependence of the cut was modeled with a phenomenological linear function, after we observed an excess broadening of the recorded light yield with respect to the photon statistics model discussed in \cite{Armengaud:2020}. The median excess width of 32\, eV ($\sim40\%$) at \qbb is associated with an under-sampling of the faster LD pulses and is presently under further investigation. A modified photon statistics model is also considered as a systematic in the limit setting.

\begin{figure}[bht]
    \centering
    \includegraphics[width=0.49\textwidth]{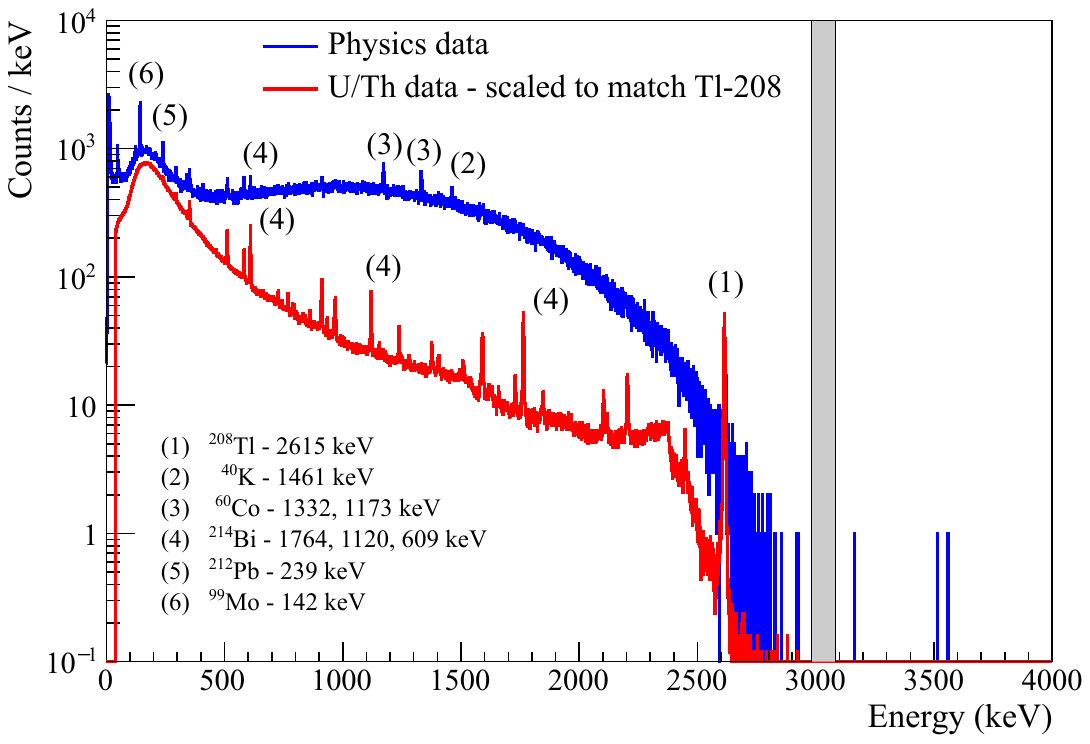}
    \caption{Physics spectrum (blue) for 2.16\,\ky of data and calibration spectrum (red) scaled to match the 2615\,keV counts from \Tl. A $\pm\, 50 \, $keV region around \qbb has been blinded (gray). 
    } 
    \label{fig:physics_spectrum}
\end{figure}

The resulting physics spectrum summed over 19 LMO detectors and the entire data taking period is shown in Fig.~\ref{fig:physics_spectrum} in blue. 
Due to the short \nnbb half-life (high rate) of \Mo \cite{Arnold:2019,Armengaud:2020b} a smooth \nnbb component dominates the spectrum from 0.5 to 3\,MeV. 
A limited set of \ga peaks remains visible, most notably \Tl, \K, \Co and an activation \ga peak from $^{99}$Mo, present for a short time after a neutron irradiation of the detectors \cite{Arnaud:2020}. For more details we refer to a prior characterization of the backgrounds in the EDELWEISS facility~\cite{Armengaud:2017b}.

We optimize the \onbb search for a Poisson counting process in the low background regime. We consider detector and data-set based ($19\times7$) energy resolutions, a preliminary estimate of our background index and an exposure of ~2.8\,\ky as we intend to replicate the present analysis for the full exposure of the now completed CUPID-Mo experiment.

The most representative $\gamma$-peak for the \Mo region of interest (ROI) with sufficient statistics to extract detector and data-set based resolutions is the 2615 keV line from \Tl in calibration data. 
We perform a simultaneous unbinned extended maximum likelihood fit of this peak with individual parameters for the detector resolutions, peak amplitudes and position and with common parameters for the peak-background ratio \cite{Schmidt:2020}. We then project these resolutions with a global scaling factor $s = \sigma_{phys}(3034\, \mathrm{keV}) / \sigma_{cal}(2615\, \mathrm{keV})$ (common to all data-sets and detectors) to \qbb.
In addition to the method described in \cite{Schmidt:2020} we extract $s$ from a polynomial fit of the global $\gamma$ peaks in background/calibration data \cite{Adams2020, Alduino:2018}. We adopt the scaling factor from this latter method as a conservative choice predicting a 0.2\% worse resolution of $(7.6 \pm 0.7 \mathrm{(stat.)} \pm 0.2 \mathrm{(syst.)}) \, \mathrm{keV}$  FWHM at \qbb for the overall data taking. 
 The noted systematic uncertainty of 2\% is due to pile-up related non-Gaussian tails in calibration data that affect the calibration resolution estimates through the PCA cut.

The background index has been evaluated from the still blinded data with a phenomenological fit model that contains an exponential to approximate both the high energy part of the \nnbb spectrum as well as tails from U/Th contaminants in the setup, and a constant as a conservative estimate for the coincident detection of two \nnbb events in the same crystal, remaining un-vetoed muon events and close contamination from the high energy beta decays in the natural U/Th chains. The result of an unbinned extended maximum likelihood fit is strongly dependent on the low-energy and high-energy limit of the fit range. 
\edit{ For a fit with the low-energy limit varied from 2.65--2.9\,MeV and the high-energy limit from the upper end of the blinded region to 4\,MeV
 we obtain a background index of $2\times10^{-3}\,$\ckky to $6\times 10^{-3}\,$\ckky in a 10\,keV window around \qbb . }
Considering the large remaining uncertainty we round the background index for the ROI optimization to ${b = 5\times 10^{-3}\,}$\ckky. We model the background as locally flat, consider detector and data-set based resolutions, and simulate the \onbb peak containment in our Geant4 Monte Carlo model. 
\edit{As this background index is both poorly constrained and indicative of a most probably background free \onbb search, we select the ROI maximizing the mean limit setting sensitivity for a Poisson process with zero background \[\overline{S_{90}} =\sum_{i=0}^{ \infty} P(i,b,\Delta E_{\mathrm{ROI}})\cdot S_{90}(i) \]
with the sum running over the product of the Poisson probability $P(i,b,\Delta E_{\mathrm{ROI}})$ of obtaining $i$ events for an ROI with width $\Delta E_{\mathrm{ROI}}$ and a background index $b$ times the expected classical 90\% confidence exclusion limit $S_{90}(i)$.}
We transfer this maximization
 from the optimization of the energy range for a peak search in 19 (detectors) times 7 (data-sets) to the optimization of a single parameter by splitting the simulated smeared \onbb peaks into 0.1 keV bins and ranking each bin associated with a triplet (detector, data-set, energy) in Signal-Background (B/S) likelihood space. 
 The optimal cutoff parameter (B/S)$_{cutoff} $ results in a ROI that is on average (exposure weighted) 17.9 keV wide. 
 It has a mean signal containment of $75.8\%$ with a spread of $\pm 1.0\%$. The ROI width corresponds to an average 2.7$\,\sigma$ Gaussian coverage with the loss of \onbb decay events in the full energy peak dominated by events with energy loss from Bremsstrahlung and electron escape close to the surface of the crystals. The optimization exhibits only a mild dependence on the background index or the knowledge of the resolution, with the overall containment changing by $\pm0.7\%$ for a $50\%$ change in $b$ ($2.2\,$keV wider, $1.5\,$keV narrower ROI). We truncate the computation of the mean limit setting sensitivity after the first three terms as the probability of 3 or more background events is negligible for the considered ROI.

\begin{figure}[tb]
    \centering
    \includegraphics[width=0.49\textwidth]{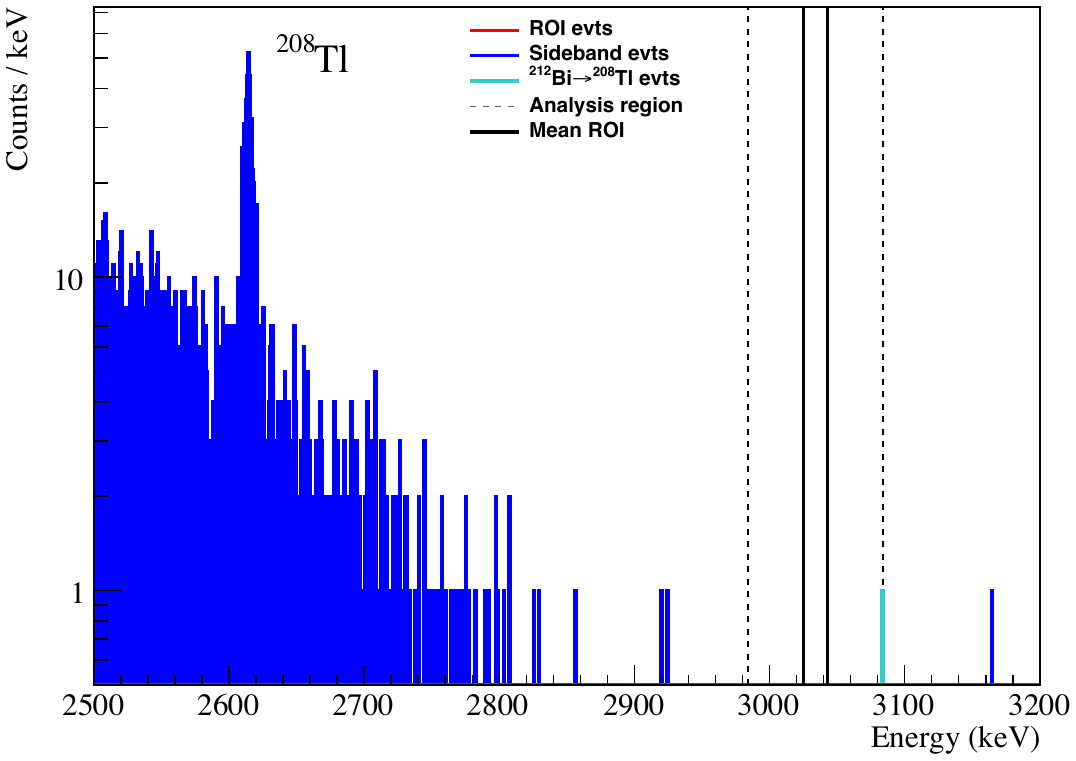}
    \caption{Physics spectrum for 2.16\,\ky of data after unblinding. No event is observed in the detector and data-set based ROI. \edit{A single event, highlighted in cyan has been observed in the analysis region. In a further refinement of the analysis it was identified as a $\beta$-candidate out of the $^{212}$Bi$\rightarrow^{208}$Tl$\rightarrow ^{208}$Pb part of the natural decay chain (see text).} For visualization the exposure weighted mean ROI for \onbb decay (17.9\,keV wide) has been indicated with solid black lines. }
    \label{fig:unblinding}
\end{figure}

As the discussed Poisson sensitivity is by construction only applicable for limit setting we implemented a binned likelihood analysis instead to either extract the final limit or a potential signal on the rate of \onbb events.   
This analysis is built on the Bayesian Analysis Toolkit (BAT) \cite{Caldwell:2009} and considers both the signal region as well as the sidebands of our 100 keV wide blinded region. The likelihood function \[L =\prod_{i=1}^{3} \frac{e^{\lambda_i} \lambda_i^{n_i}}{n_i !},\]  is the product over three Poisson terms for the two sidebands and the signal ROI with observed events $n_i$ and expected events $\lambda_i$. The mean number of expected events $\lambda_i$ is computed considering the phenomenological background model described above and a Gaussian signal contribution in which we leave the strength of the signal and flat background component free by using uninformative flat priors. After defining all analysis steps we unblind and obtain the spectrum in 
 Fig. \ref{fig:unblinding}. We observe no event in the signal region and a single event (cyan) in the right-hand side region.
The corresponding marginalized posterior distribution for the number of signal events has a most probable value of zero with an upper limit  of 2.4  events at 90\%~C.I., resulting in a half-life limit for \onbb decay in \Mo of  $T_{1/2}^{0\nu} > 1.4 \times 10^{24}\, \mathrm{yr}\quad \mathrm{(90\%~C.I.)}.$ 
The posterior for the flat background is non-zero with a 1$\,\sigma$ interval of $3\substack{+7 \\ -3} \times 10^{-3}\,$\ckky and
the posterior distributions for the parameters of the exponential are compatible with priors from a fit of the \nnbb spectrum in the 2650--2980\,keV interval. 
We repeat the same fit for the approximation of a Gaussian signal with a locally flat background over the 100 keV analysis region. 
The limit on \onbb decay of \Mo is unaffected.

\begin{table}[hb]
\caption{\label{tab:Nuisance}%
Nuisance parameters included in the analysis and their implementation with Flat or Gaussian prior in the Bayesian fit. Parameters 2 and 4 are multiplicative scaling factors instead of absolute uncertainties, see text for details. }
\begin{ruledtabular}
\begin{tabular}{lccc}
\textrm{Systematic}& \textrm{Index} &
\textrm{Value}&
\textrm{Prior}\\
\colrule
\onbb detector response & 1& 0.95--1.00 & Flat \\
\onbb containment MC & 2&$1.000\pm0.015$ & Gauss \\
Analysis efficiency $^{\mathrm{a}}$ & 3 &$0.906\pm0.004$ & Gauss \\
Light yield selection \footnote{Data-set dependent; exposure weighted mean value presented.} & 4& 0.998--1.008 & Flat \\
Isotopic enrichment & 5&$0.966\pm 0.002$ & Gauss \\
\end{tabular}
\end{ruledtabular}
\end{table}
The nuisance parameters considered in this limit are summarized in Table \ref{tab:Nuisance}. Uncertainties on the detector response, in particular the energy scale and resolutions, are included in the simulation of \onbb events. They are hence covered in the resulting containment in the optimized central ROI on a detector and data-set basis and not considered independently. The only remaining uncertainty for the detector response (Index 1, Table \ref{tab:Nuisance}) is based on a potential non-gaussianity of the \onbb peak. In this analysis we estimate this contribution based on the shape of the 2615 keV calibration peak. We observe evidence for non-gaussian tails, which are dominated by unrejected pile-up events caused by the high trigger-rate in calibration data. We set a conservative systematic on the containment reduction of up to 5\%. 
The second nuisance parameter on the containment (Index 2) accounts for the Geant4 modeling uncertainty of Bremsstrahlung events. 
Reported accuracies for the Geant4 Bremsstrahlung production of a few MeV electrons in thick targets \cite{Faddegon:2008, Pandola:2015} of $\sim$10\% result in a systematic uncertainty in the overall containment of the \onbb signal in the optimized ROI of $(75.8\pm1.1)\% $ for our crystal geometry which is applied as a single common multiplicative factor of $1.000\pm0.015$ in the limit setting. 
The inclusion of the analysis efficiency \[\epsilon = (90.6 \pm 0.4 \,\mathrm{(stat.)}\, \substack{ +0.8\\ -0.2}\,\mathrm{(syst.)} )\%\] is split into two parts. For the evaluation of the mean value and its statistical uncertainty (Index 3), we make use of the two independent signals in the LDs and LMOs to evaluate cut efficiencies on a clean sample of signal events in the  1.3\,MeV to 2\,MeV \nnbb spectrum or from the \Po peak \cite{Schmidt:2020}. Energy independent cuts are evaluated directly from the ratio between passed and total events with binomial uncertainty. The pulse shape analysis efficiency is extracted from a linear fit extrapolated to \qbb in order to account for the energy dependence in the reconstruction error. 
\edit{
The systematic uncertainty associated with the excess broadening of the light yield cut has been evaluated with a set of pseudo-experiments considering the linear and modified photon statistics model introduced before. It is reflected in our limit setting as a multiplicative factor with uniform prior in 0.998--1.008 (Index 4).}
Lastly, we include a subdominant uncertainty in the enrichment and number of \Mo atoms of 0.2\% (Index 5).

\edit{We further refined our analysis after unblinding, implementing a cut designed to reject high energy $\beta$ events from the $^{212}$Bi$\xrightarrow{\alpha}$$^{208}${Tl}$\xrightarrow{\beta}$$^{208}$Pb branch in the thorium chain ($T_{1/2}^{^{208}\textrm{Tl}}=183\,$s, 5\,MeV $Q$-value). 
Similar to previous analyses with scintillating bolometers \cite{Pirro:2005ar, Azzolini2018b, Azzolini:2019nmi} we tag  $^{212}$Bi $\alpha$ candidates with energies in the 6.0--6.3 MeV range, and veto any decay in the same crystal in a 10 half-life period (1832\,s). 
This cut has a negligible impact on the life-time (0.02\%) accidentally rejecting 2:10000 events, but it does reject the event close to the ROI in cyan in Fig.~2.
The energy of the preceding \al candidate is consistent with the $Q$-value of $^{212}$Bi within 10 keV and the time difference between the events is 113\,s.}
\edit{We report a final \onbb limit that is 1.3\% stronger and rounds to  \[{T_{1/2}^{0\nu} > 1.5 \times 10^{24}\,\mathrm{yr} ~ \mathrm{(90\%~C.I.)}}.\]}  \noindent The posterior for the flat background of the bayesian fit in this case is peaked at zero with a 90\%~C.I. of $1.1 \times 10^{-2} \,$\ckky. 

 We interpret the obtained half-life limit in the framework of light Majorana neutrino exchange using ${g_A=1.27}$, phase space factors from 
 \cite{Kotila:2012, Mirea:2015}
 and nuclear matrix element calculations from \cite{Rath:2013,Simkovic:2013,Vaquero:2013,Barea:2015,Hyvarinen:2015,Song:2017,Simkovic:2018,Rath:2019}. The resulting limit on the effective Majorana neutrino mass of $\langle m_{\beta\beta} \rangle < $(0.31--0.54)$\,$eV is the fourth most stringent limit world wide, obtained with a modest \Mo exposure of 1.17\,\ky. 
 It is the leading constraint for \Mo, exceeding the previous best limit from NEMO-3 \cite{Arnold:2015} by $30\%$ with almost 30 times lower \Mo exposure.
The technology of CUPID-Mo has proven that it can be operated reliably, reaches high efficiency for \onbb search of $68.6\%$ (containment $\times$ analysis efficiency) and a resolution of ${0.11\%~(1 \sigma)}$ at \qbb. 
The present analysis strengthens the projection of the CUPID sensitivity \cite{CUPIDInterestGroup:2019inu}, by demonstrating a detailed understanding of the \onbb ROI and confirming key assumptions like the efficiency of \enrlmo based cryogenic scintillating bolometers.  
Extremely low U/Th contamination levels in the \lmo crystals reported in \cite{Schmidt:2020} surpass the requirements for CUPID \cite{CUPIDInterestGroup:2019inu}, and an efficient alpha separation has been demonstrated both in cylindrical \cite{Armengaud:2017, Armengaud:2020} and recently also in cubic \lmo detectors \cite{Armatol2021b,Armatol2021a}. 
The preliminary estimate of the background in the ROI at the few $10^{-3} \,$\ckky level in CUPID-Mo, obtained in an experimental setup that was not designed for a \onbb search, is encouraging and supports our believe that a $10^{-4} \,$\ckky background level for CUPID \cite{CUPIDInterestGroup:2019inu} seems feasible.

\edit{Further analyses from CUPID-Mo will be focused on precisely reconstructing remaining backgrounds, comparing to the best reported background index for a bolometric \onbb search $(3.5\substack{+1 \\ -0.9}\times10^{-3}\,$\ckky) in CUPID-0 \cite{Azzolini:2019tta,Azzolini:2019nmi} and to optimally design and use the technology of the CUPID-Mo experiment in CUPID \cite{CUPIDInterestGroup:2019inu}.}

The help of the technical staff of the Laboratoire Souterrain de
Modane and of the other participant laboratories is gratefully
acknowledged. This work has been partially performed in the
framework of the LUMINEU program, funded by ANR (France). This
work was also supported by CEA and IN2P3, by P2IO LabEx with the BSM-Nu project, by Istituto Nazionale di Fisica Nucleare, the Russian Science Foundation
(Russia), the National Research Foundation (Ukraine), the US
Department of Energy (DOE) Office of Science, the DOE Office of
Science, Office of Nuclear Physics, the National Science
Foundation (USA), by the France-Berkeley fund, the MISTI-France
fund, and by the Chateaubriand Fellowship of the Office for
Science \& Technology of the Embassy of France in the United
States. Individuals have received support from the National
Academy of Sciences of Ukraine and P2IO LabEx managed by the ANR
(France). This work  makes  use  of  the  DIANA  data  analysis  software which  has  been  developed  by  the  Cuoricino,  CUORE, LUCIFER, and CUPID-0 Collaborations.


\bibliography{Biblio.bib}
\clearpage








\end{document}